# Downtrack response of differential reader for high density magnetic recording

Li Wang*, Guchang Han

*Data Storage Institute, Agency for Science, Technology and Research (A*STAR),*

*5 Engineering Drive 1, 117608, Singapore*

The downtrack responses of a differential reader to various magnetic patterns have been investigated micromagnetically. The differential signals can well discriminate the "0" and "1" readback bits and the waveforms obtained resemble the counterparts for longitudinal media or for perpendicular media after differentiation. Pulse shapes are found Gaussian. With decreasing head media spacing, free layer thickness, and gap length, $PW_{50}$ roughly linearly decreases and the maximum signal exponentially increases. These properties, together with small inter-symbol interference, are essential for future high density magnetic recording.

*Electronic mail: wang_li@dsi.a-star.edu.sg.

Maintaining of growth of HDD areal densities and fending off the probable competitions arisen from SSDs[1] demand more advancements and understandings in recording system including read heads. AMR-based differential readers[2-4] (DRs) can reduce common-mode-rejection noise while GMR-based DRs[5] are shield free because of the replacement of shields by gap layer to define linear density. Almost all the works so far for GMR-based DRs are in a conceptual stage[5,6]; micromagnetic modeling of differential readers (responded to media) is lacking and the corresponding properties are unclear. Thus, it would be promptly to test the conceptual claims and evaluate the read performances of the GMR-based DRs. Here perpendicular media bits with various magnetization configurations are used in a finite-element micromagnetic model[7,8] to test the differential readbacks of the "0" and "1" bits. Note that an explicit formulation of the GMR-based DR signals is also lacking and such formulation is valuable for clarifications of the results obtained.

Our simulations firstly indicate that differential readers can well discriminate the "0" and "1" readback bits and the corresponding responses are similar to the conventional playbacks from longitudinal media[9] or from perpendicular media after differentiation[10]. Furthermore, differential readers are found to have small inter-symbol interference (ISI) which are suitable for shield-free and for high density magnetic recording; here, the scenario of "shield-free" is different from Ref. 6 where "shield-free" refers to the replacement of shields by gap layer. Finally, we have studied properties of pulse shape, $PW_{50}$, and the maximum playback signal; these studies further reveal the good compatibility, sensitivity and scalability of differential readers.

Fig. 1 illustrates our modeling system with the definitions of GMR and DR signals. The very essential components of GMR-based DRs (hereafter simply termed as DRs) are two free layers [FLs; with parallel (or antiparallel) magnetizations and separated by a gap layer] with two reference layers (RLs) of opposite magnetizations. The downtrack responses are computed by solving the Landau-Lifshitz-Gilbert (LLG) equation[7]. Here, typical simulations parameters are (length unit: nm; dimensions are defined in the order: crosstrack x downtrack x vertical): FL dimensions: 10x2x8; hard bias (HB) dimensions[8]: 10x13.332x8; bit dimensions: 12x12x2; Gap length (GL) = 4 [defined as the distance between the inner surfaces of the two FLs, see Fig. 1(b)], FL-to-HB gap = 5, magnetic spacing = 3 (head media spacing; defined as the distance between the bottom of FLs and the top of the bits). Saturation magnetizations: $M_s^{\text{media}} = 500$ emu/cc, $M_s^{\text{head}} = 800$ emu/cc; anisotropy constants: $K^{\text{media}} = 1\text{x}10^6$ J/m$^3$, $K^{\text{head}} = 5\text{x}10^2$ J/m$^3$. For simplicity, exchange constants are set at 1.3x10$^{-11}$ J/m and damping constants at 0.2; the $M_s$ and $K$ parameters of hard biases are assumed similar to those for media[8]. Typical mesh is 2x2x2 nm$^3$ and the errors for smaller meshes are small.

The GMR (or TMR) signal[11], is $V(\theta) = IR_{\uparrow\uparrow}[1 + 0.5(1 - \cos\theta)\Delta R / R_{\uparrow\uparrow}]$ where $\theta$ is the angle(s) between the FL and RL magnetizations [see Fig. 1(c)] and $\Delta R = R_{\uparrow\downarrow} - R_{\uparrow\uparrow}$, with $R_{\uparrow\downarrow}$ and $R_{\uparrow\uparrow}$ being the respective resistances at 180$^\circ$ and 0$^\circ$. Here we use $R_{\uparrow\uparrow}$ = 50 Ω, $\Delta R$ = 5 Ω, and $I$ = 0.8 mA. The signal for usual readers is (omitting the efficiency constant) $\Delta V = V(\theta) - V(90) = -0.5I\Delta R\cos\theta$. Hence:

$$\Delta V = \begin{cases} > 0, \text{ when } \theta > 90, & 1(a) \\ < 0, \text{ when } \theta < 90, & 1(b) \\ = 0, \text{ when } \theta = 90. & 1(c) \end{cases}$$

The signal for differential readers is the summation of two GMR readings with opposite reference layers [see Fig. 1(d)], which is (assuming $|\Delta V_L| = |\Delta V_R|$):

$$\begin{aligned} V &= \Delta V_L + \Delta V_R = \pm|\Delta V_L| \pm |\Delta V_R| \\ &= \begin{cases} 2|\Delta V_L|, & \Delta V_L > 0,\ \Delta V_R > 0, \quad 2(a) \\ -2|\Delta V_L|, & \Delta V_L < 0,\ \Delta V_R < 0, \quad 2(b) \\ +|\Delta V_L| - |\Delta V_R| = 0, & \Delta V_L > 0, \Delta V_R < 0, \quad 2(c) \\ -|\Delta V_L| + |\Delta V_R| = 0, & \Delta V_L < 0, \Delta V_R > 0, \quad 2(d) \\ 0, & \Delta V_L = 0, \Delta V_R = 0. \quad 2(e) \end{cases} \end{aligned}$$

This equation represents differential signals.

Fig. 2 plots the DR signals and responses for various magnetic patterns. We have noticed that: (I) For the signal at point A (offset -24 nm) in Fig. 2(a), the upward fields clearly render $\theta_L > 90$ , $\theta_R < 90$ [see case A in Fig. 2(b)]. Hence, $\Delta V_L > 0$ , $\Delta V_R < 0$ (Eq. 1). According to Eq. 2(c), the DR signal is 0. The same analysis applies to point B (offset - 18 nm, where the two FLs fall within the range of one media bit) and other points since the field directions remain upward. (II) At point A in Fig. 2(c), the signal is negatively minimum. This is associated with $\theta_L < 90$ , $\theta_R < 90$ in case A in Fig. 2(d), thus, $\Delta V_L < 0$ , $\Delta V_R < 0$ ; according to Eq. 2(b), the DR signal is negative (minimum). At point B, the two FLs fall within the range of one bit with downward field direction. Hence, $\theta_L < 90$ , $\theta_R > 90$ [see case B in Fig. 2(d)] and $\Delta V_L < 0$ , $\Delta V_R > 0$ . According to Eq. 2(d), the DR signal is then 0. For case C, the situation is opposite to case A and the DR signal (positively maximum) is determined by Eq. 2(a). For case D, the situation is opposite to case B and the DR signal is determined by Eq. 2(c). (III) The responses at points A, B and D in Fig. 2(e) are the same as case A in Fig. 2(b), case A in Fig. 2(d), and case C in Fig. 2(d), respectively. At point C in Fig. 2(e), the signal is zero and in this case, $\theta_L < 90$ , $\theta_R > 90$, corresponding to $\Delta V_L < 0$, $\Delta V_R > 0$ and thus $\Delta V = 0$ [see Eq. 2(d)].

Fig. 2 indicates that DRs can well discriminate the readback bits “0” and “1”. The waveforms obtained are reasonable and particularly, the one in Fig. 2(e) is similar to the experimental observations from conventional readers responded to longitudinal media[9] or to perpendicular media after differenation[10]. This similarity enables DRs to benefit from perpendicular magnetic recording (PMR) and meanwhile avoid the differentiation (which may largely amplify electronic noise); moreover, the well-established signal processing for longitudinal magnetic recording (LMR) may be re-used[6]. It should be stressed that the flat shape in Fig. 2(a) and the periodicities and the equal response amplitudes observed in Figs. 2(c) and (e) imply that DRs mainly respond to local media fields and are resilient to global fields emanated from other parts of media. Hence, DRs would possess less ISI and thus can be shield free and can facilitate higher linear density. Here, "shield free" is quite different from the scenario[6] that shields in DRs are unnecessary just because in DRs the shield-to-shield spacing is replaced by gap length to define linear density.

The dipulse in Fig. 2(e) demonstrates isolated-like behaviors [plateaus approching 0 at points A, C, E]. Thus we shall use the underlying media pattern ↑↑↓↓↑↑ for more investigations. Fig. 3 summarizes the results calculated for the pulse shapes, $PW_{50}$, and the maximum readback signal ($V_{\mathrm{peak}}$). For large magnetic spacing, thick free layer, and large gap length, we have virtually observed an asymmetry effect: the signal for the "0" bit at 24 nm is somewhat larger than the counterpart at 0 nm. This is in contrast to the equal *zero* amplitudes at points C and E in Fig. 2(e). The cause for this asymmetry effect is that the DRs with large gap length or thick free layer or large magnetic spacing will become more sensitive to the bit nearby environments (i.e., more ISI) and thus shields may be needed; Contrarily, the DRs with small magnetic spacing, thin free layer, and small gap length, which are judicious choices for future high density magnetic recording, will be of less asymmetry effect (ISI), shields-free, and with simplified fabrication process. The asymmetry effect in Figs. 3(a) and (c) is actually not obvious. However, for Figs. 3(e) and (f), we have used the pattern ↑↑↓↓↑↑↓↓ to mitigate large asymmetry effect at large GLs observed for ↑↑↓↓↑↑; the HB thicknesses are also increased so that a large and uniform stabilization can be provided in the case of large GLs.

It can be found from Fig. 3 that: (I) pulse shapes of the transitions can be approximated by Gaussian functions with skirts of zero tails (except the deviations at *very large* GLs). One can verify that Gaussian-type behaviors also hold for the PMR signal (the derivative of the error function in Eq. 22 in Ref. 12 is Gaussian) and for the LMR signal (the shape in Fig. 8 in Ref. 13 can be fitted by Gaussian); (II) $PW_{50}$ increases nearly linearly with increasing magnetic spacing, free layer thickness, and gap length. This linearity is a good approximation to and reminiscence of the root-square relations for $T_{50}$ in PMR[12] and for $PW_{50}$ in LMR[14]. (III) The maximum signals decrease exponentially with increasing magnetic spacing, free layer thickness, and gap length (not too large). When the GL is relatively large, the maximum signal actually increases [see Fig. 3(f)]. The linear and nonlinear behaviors in Fig. 3(b) are similar to the LMR case[13]. The exponential decay in Fig. 3(b) is ascribed to spacing loss[15]. To explain the exponential-decay readbacks in Figs. 3(d) and (f), we expected and did find that media field exponentially decays away from a transition, see, e.g., Fig. 4(a), where the Ansys data (see also Ref. 15) for the transition roughly follow the expression for a step transition[16]:

$$B_y = \frac{u_0 M_s}{\pi}\left[\tan^{-1}\left(\frac{x}{y+\delta/2}\right) - \tan^{-1}\left(\frac{x}{y-\delta/2}\right)\right], \qquad (3)$$

with $\delta$ being media thickness. However, we further realized that: (I) The absolute value of the demagnetizing field along the stripe height direction[17] (which is parallel to media field direction) increases exponentially with decreasing FL thickness [cf. Fig. 4(b)]; such improved sensitivity should most contribute to exponential behavior in Fig. 3(d) as we have found that both $V_{\text{peak}}$ and $-D_{SH}$ decrease *monotonically* with FL thickness while the downtrack $B_y$ profile is *non-monotonic* [compare also the data (○) in Figs. 3(d), 4(b) to 4(a)]. (II) The increase of $V_{\text{peak}}$ in Fig. 3(f) for larger GLs is correlated to the media fields averaged within the FLs while the sharp drop of $V_{\text{peak}}$ at smaller GLs is related to the interactions among the two FLs and the media bits. More efforts[15] are needed to understand these interactions (and other topics such as effects of media thicknesses and patterns).

In summary, differential responses to perpendicular bits with different configurations have been investigated micromagnetically. Differential readers can well discriminate the readback bits and typical pulse shapes are Gaussian. Appropriate designs of gap length, free layer, and magnetic spacing can lead to high playback signals (thus high sensitivity) with small $PW_{50}$ (thus good scalability). These properties, combined with other merits such as good compatibility, small inter-symbol interference, and easy fabrication, make differential readers a viable reading candidate for future high density magnetic recording.

We thank Dr. Z. M. Yuan for discussion on $PW_{50}$, Mr. Y. K. Yeo for help on Ansys, Dr. T. Coughlin for communication on media parameters, and Dr. B. Liu for useful remarks.

## LIST OF REFERENCES

## LIST OF FIGURE CAPTIONS

FIG. 1. Perspective (a) and top (b) views of the simulation system: a differential reader with media bits and hard biases. Gap length (GL) and bit length (BL) are indicated in (b). (c) & (d) show schematically the definitions of GMR and DR signals, respectively. The reference layers (RLs) for the two free layers ($FL_L$ and $FL_R$) have opposite magnetizations, see (d).

FIG. 2. Waveforms of differential readers for readback bits of “00000” (a), “11111” (c), and “01010” (e). The corresponding media patterns, with typical DR responses, are illustrated in (b), (d) & (f), respectively. The downtrack postion (offset) is the displacement of the head gap center against the middle location of the six media bits. The peaks in (c) and (e) indicate the “1” bits for the transtions while the plateaus-like behaviors at points A, C, E in (e) indicate the “0” bits representing non-transtions.

FIG. 3. Pulse shapes are Gaussian as exemplifed by several values (unit nm) of magnetic spacing, free layer thickness, and gap legnth, see (a), (c) & (e) with the exception at large GLs. The variations of $PW_{50}$ and maximum signals ($V_{peak}$) with magnetic spacing, free layer thickness, and gap length are shown in (b), (d) & (f).

FIG. 4. (a) Calculated $B_y$ profile near a transition (○), as compared to Eq. (3). (b) Negative demagnetizing factors of a free layer with stripe height (SH) 8 nm, track width (TW) 10 nm, and thickness (TH) 2 nm. The solid lines represent exponential fits.

Fig. 1

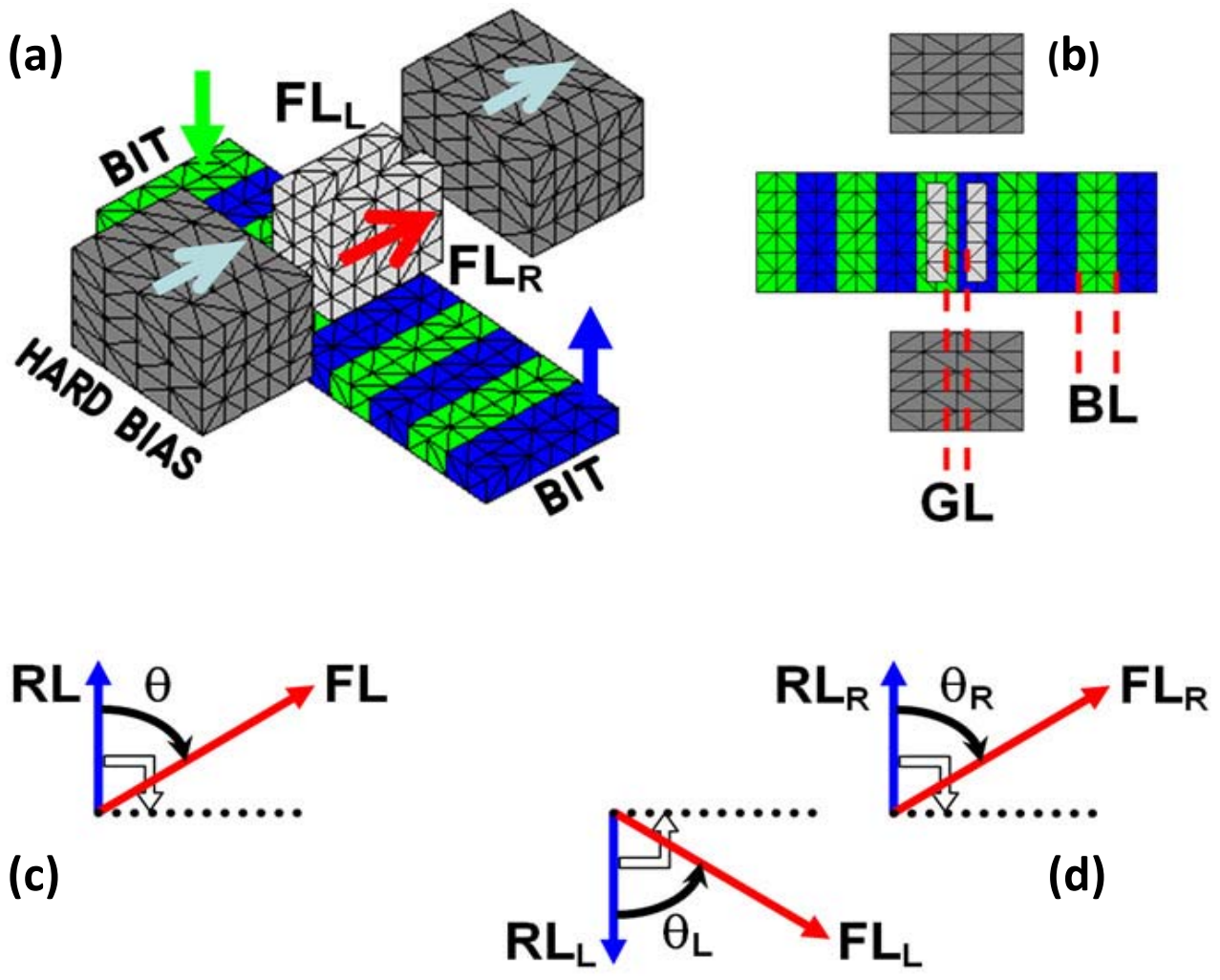

Fig. 2

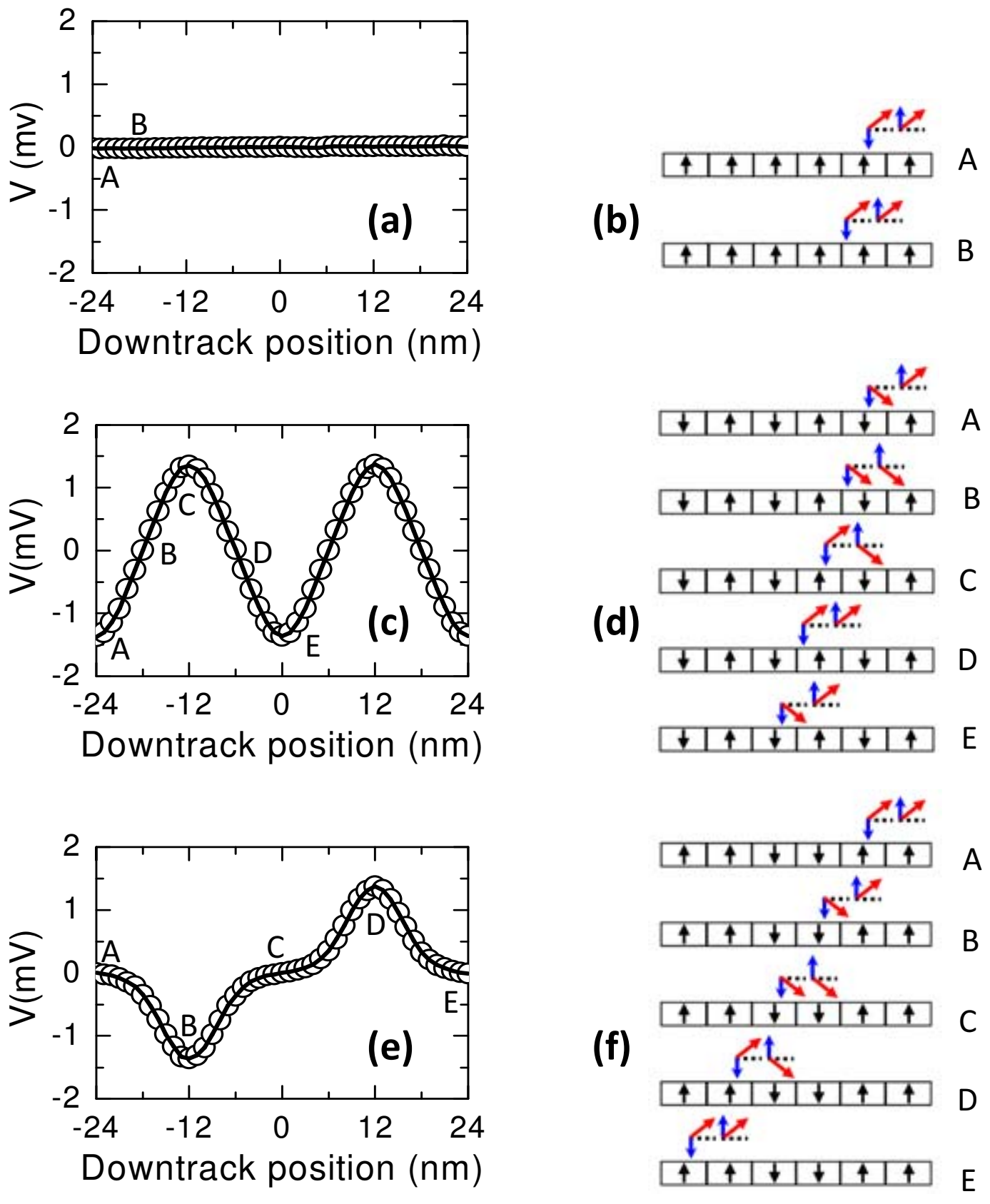

Fig. 3

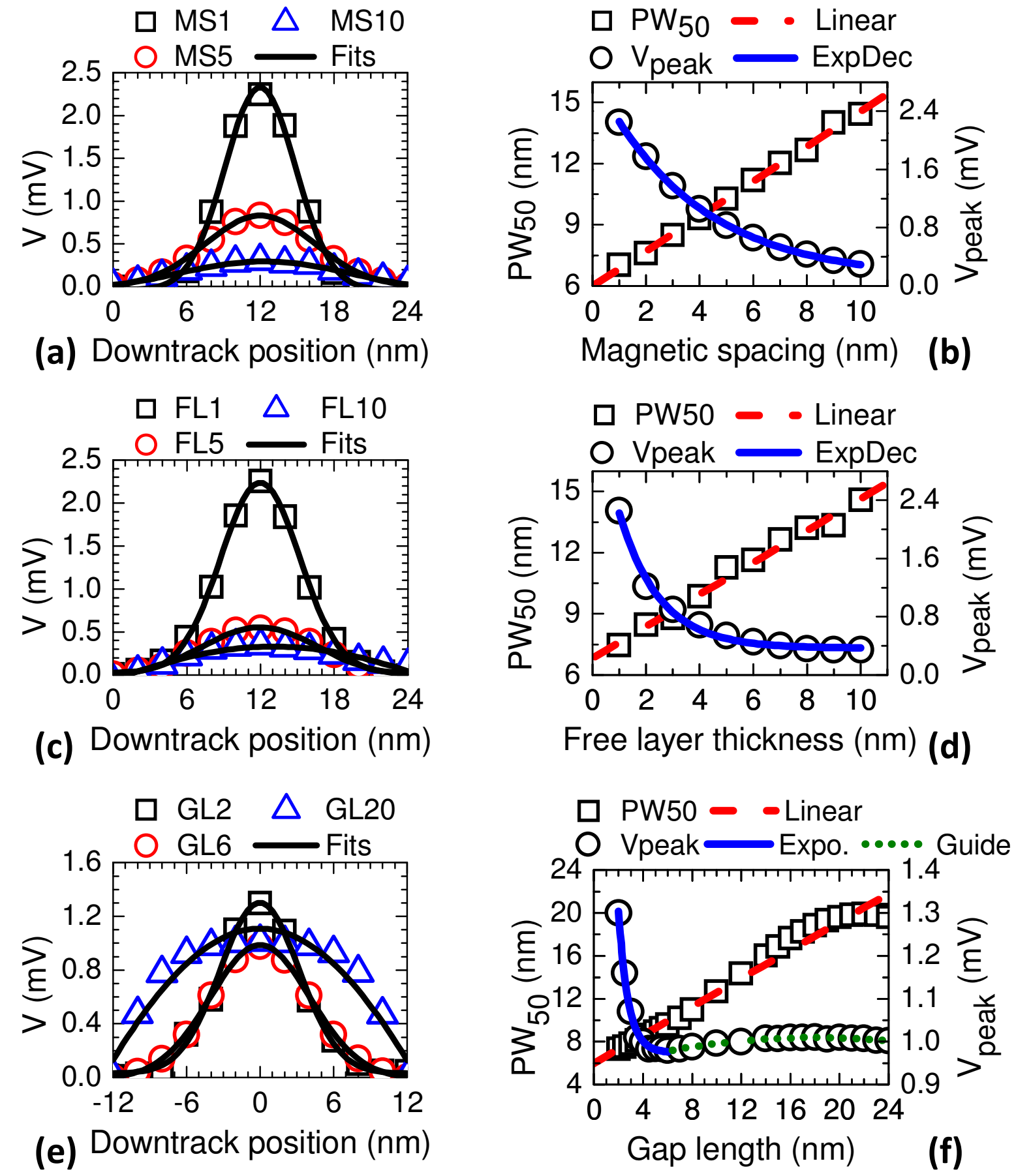

Fig. 4

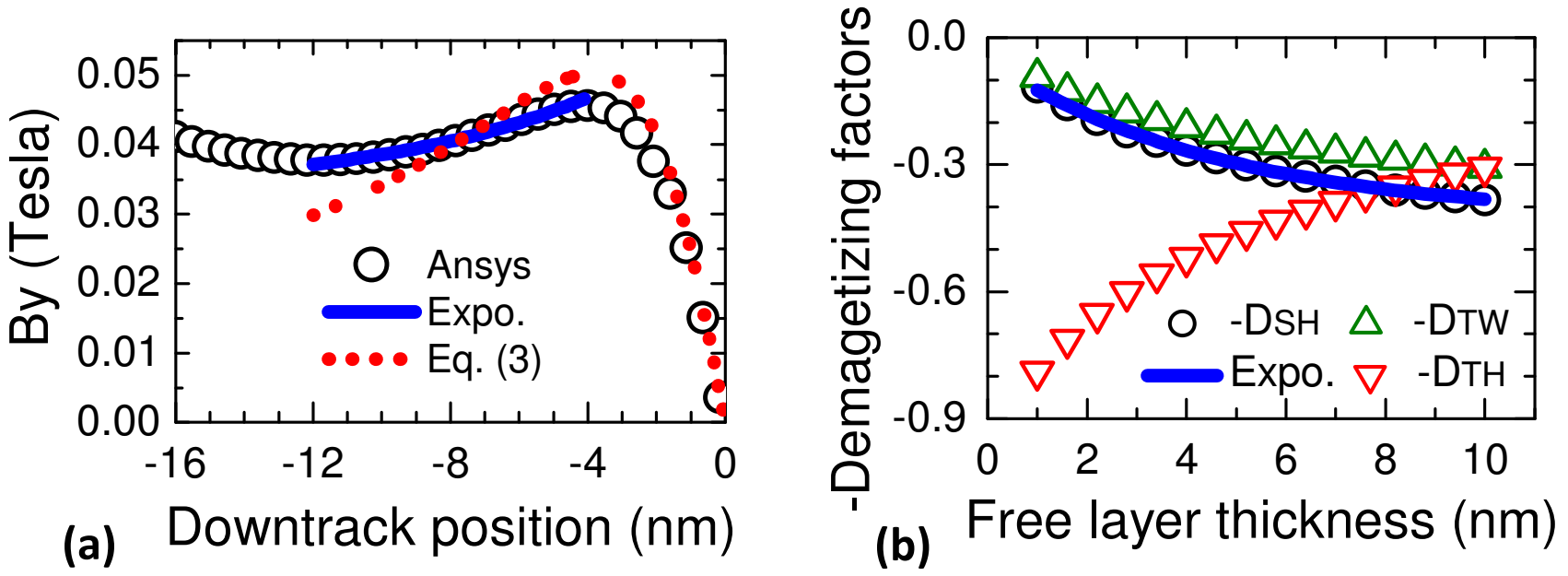